\documentclass[pdftex,twocolumn,epjc3]{svjour3}          
\RequirePackage[T1]{fontenc}

\smartqed  
\usepackage{latexsym}
\usepackage{url}
\usepackage{amsmath}
\usepackage{amsfonts}
\usepackage{setspace}

\RequirePackage{graphicx}
\RequirePackage{mathptmx}      
\RequirePackage{flushend}
\RequirePackage[numbers,sort&compress]{natbib}
\RequirePackage[colorlinks,citecolor=blue,urlcolor=blue,linkcolor=blue]{hyperref}

\journalname{Eur. Phys. J. C}

\begin{document}

\title{Evolution of perturbed dynamical systems: analytical computation with time independent accuracy}

\author{A.V.~Gurzadyan\thanksref{addr1,e1}
        \and
        A.A.~Kocharyan\thanksref{addr2} 
}

\thankstext{e1}{e-mail: mail@xxx.xx}

\institute{Department of Mathematics and Mathematical Modelling, Russian-Armenian (Slavonic) University, 
					Yerevan, Armenia\label{addr1}
          \and
          School of Mathematical Sciences, Monash University, Clayton, Australia\label{addr2}
}

\date{Received: date / Accepted: date}

\maketitle

\begin{abstract}
An analytical method for investigation of the evolution of dynamical systems {\it with independent on time accuracy}
is developed for perturbed Hamiltonian systems. The error-free estimation using of computer algebra enables the 
application of the method to complex multi-dimensional Hamiltonian and dissipative systems. It also opens principal 
opportunities for the qualitative study of chaotic trajectories. The performance of the method is demonstrated on 
perturbed two-oscillator systems. It can be applied to various non-linear physical and astrophysical systems, 
e.g. to the long-term planetary dynamics.
\end{abstract}

\section{Introduction}

The high accuracy investigation of the evolution of dynamical systems is a principal task in various astrophysical and physical problems,
e.g. trajectories of artificial satellites testing the General Relativity \cite{C}, orbits and precession of planets of the Solar system \cite{L,L1}, galactic dynamics and gravitating N-body problem  \cite{A}, etc. Therefore, development of more advanced methods, especially of those enabling non-iterative, and hence with negligible error storage at each iteration, for description of evolution of dynamical systems or broadening the classes of systems to which they are applicable, are of particular interest. 

Among the principal issues is the numerical treatment of systems in chaotic regimes, including the tools for revealing the chaos. For example, since, on the one hand, Lyapunov exponents are characteristics in the limit of infinite time, on the other hand, long iterative trajectory computations lead to inevitable accumulation of errors, suggested numerical procedures for n-dimensional non-linear systems (e.g. \cite{Ben}) can be applied and interpreted with limited efficiency. Hence, time error-free schemes developed even for particular models such as \cite{Ein}, can be efficient. 

Methods of theory of dynamical systems are rather efficient for numerical studies of complex systems, including chaotic, random ones  \cite{Arn,Arn2008,Arn2009}, and that approach is used below. We develop further the resolvent method suggested in \cite{GK1,GK2}. The main issues elaborated below are: 

1. Avoiding of the explicit use of Laplace and inverse Laplace transforms, and obtaining an essentially simplified evolution function representation, we broaden the classes of dynamical systems to which the method is
applicable, namely, to non-linear Hamiltonian systems with many degrees of freedom. This is crucial, particularly, for N-body codes \cite{A}.

2. The evolution function is obtained analytically i.e. with an accuracy independent of time and establishes the asymptotic and time-average characteristics of the system and, hence, enables one to deal also with chaotic trajectories. To this end, the error-free estimation of phase point (state) evolution in time allows to characterise the behavior and system's accessory at ergodic or mixing phase space regions. This is, again, a principal topic for gravitating N-body systems known to possess chaotic properties \cite{G}.  

3. The choice of the initial function, i.e. of the one not affected by perturbations, as a periodic function is justified in various physical contexts, the newly applied procedure does not involve the Fourier expansion used previously at resolvent computation. Hence, the Hamiltonian and the initial function can be represented in more convenient variables depending on the particular dynamical system and aims, e.g. unbounded and non-periodic, differing from action-angle variables.

Hamiltonian formalism introduced for dissipative systems needs specific steps, and the transformation
to even-dimensional Liouville system usually mounts additional constraints,  e.g. the dual Hamiltonian description \cite{Bat,Um} 
for damping systems requires introduction of auxiliary supplement. Based on this, the developed method allows one to follow the trajectory change of the Hamiltonian's principal term, not paying attention to the additional part. Then the reduction in \cite{GK2} is again applicable via the scheme introduced here.

The method developed below, as we show also with computations on sample systems, enables one the error-free study for broad classes of dynamical systems - of perturbed Hamiltonian and dissipative ones - and can be applied to complex physical systems.  

\section{The evolution function and the resolvent}

Hamiltonian equations for a perturbed system given by Hamiltonian
\begin{equation}
H(p,q,\beta) = H_0(p,q) + \beta V(p,q),
\end{equation}
are considered in \cite{GK1, GK2} in a unified form
\begin{equation}
\dot{x}^k=A^k(x)\ ,
\end{equation}
where $(x^1,x^2,\dots,x^{2n-1},x^{2n})=(p^1,q^1,\dots,p^n,q^n)$.

Systems (1) are the key ones in Kolmogorov-Arnold-Moser (KAM) theory stating the survival of the
perturbed tori when certain conditions are fulfilled \cite{Arn}.
 
Then for any smooth function $f(x)$ the following equation holds
\begin{equation}
\frac{d}{dt} f(x) - L_A f(x) = 0
\end{equation}
where
\begin{equation}
L_A =  A^k(x)\frac{\partial}{\partial x^k}\, .
\end{equation}

Considering $f(\cdot)$ as a function of generalized coordinate, conjugate momenta and time,
the equation describing the evolution of this function in the phase space changing according to (1), gets the form 
\begin{equation}
\frac{d}{dt} f(x,t) - L(\beta) f(x,t) = 0,\qquad x\equiv(p,q)\, .     
\end{equation}

Then, according to Hamiltonian equations
\begin{equation}
\begin{array}{l}
L(\beta) = L_0 + \beta L_1\, ,\\[5pt]
L_0 = \frac{\partial H_0}{\partial p_k}\; \frac{\partial}{\partial q^k}\, - \, 
\frac{\partial H_0}{\partial q^k}\;\frac{\partial }{\partial p_k}\, =\, \{H_0,\cdot\}\, ,\\[5pt]
L_1(V) = \frac{\partial V}{\partial p_k}\; \frac{\partial}{\partial q^k}\, - \, 
\frac{\partial V}{\partial q^k}\;\frac{\partial }{\partial p_k}\, =\, \{V,\cdot\}\,.
\end{array}
\end{equation}
The evolution equation can be changed from the canonical variables to action-angle ones 
used in \textit{KAM}-analysis to describe the phase portrait ($Tor^n\times R^n$) of the trajectories
(see \cite{Arn}, \S 50). Then
\begin{equation}
H(I,\vartheta,\beta) = H_0(I) + \beta V(I,\vartheta)
\end{equation}
and
\begin{equation}
\begin{array}{l}
L_0 = \frac{\partial H_0(I)}{\partial I_k}\; \frac{\partial}{\partial \vartheta^k}
=\omega^k(I)\; \frac{\partial}{\partial \vartheta^k}\, ,\\[5pt]
L_1(V) = \frac{\partial V}{\partial I_k}\; \frac{\partial}{\partial \vartheta^k} - 
\frac{\partial V}{\partial \vartheta^k}\; \frac{\partial}{\partial I_k}\, .
\end{array}
\end{equation}
where $\omega^k(I)=\frac{\partial H_0(I)}{\partial I_k}$.

By means of the Laplace transform 
\begin{equation}
\tilde{f}(x,s) = \mathcal{L}\left\{ f(x,t) \right\} = \int_0^{\infty} dt\, f(x,t)\, e^{-s t}, 
\end{equation}
one can rewrite equation (5) as follows:
\begin{equation}
s\, \tilde{f}(x,s) - L(\beta)\,\tilde{f}(x,s) = f(x,0)
\end{equation}
or
\begin{equation}
\begin{array}{l}
\tilde{f}(x,s) = R_s(L) g(x)\, ,\\[5pt]
g(x) \equiv f(x,0)\, , 
\end{array}
\end{equation}
where the resolvent $R_{s}(L)$ is equal to
\begin{equation}
R_{s}(L) = \left(s - L(\beta)\right)^{-1}\, .
\end{equation}

For any well defined operators $A$ and $B$ one can show the resolvent satisfies following relation
\begin{equation}
R_s(A) - R_s(B) = R_s(A)\left[A-B\right]R_s(B)\,.
\end{equation}

Then, by means of the inverse Laplace transform, the time domain of $R_{s}(\cdot)$ can be defined, e.g. for $A$
\begin{equation}
\mathcal{L}^{-1} \left\{ R_s(A) \right\}=\int_C\frac{d s}{2\pi i}\, e^{st}R_s(A)=e^{tA}\, ,
\end{equation}
where the contour of integration is 
\begin{equation}
C=\{ \sigma +i\gamma,\, \sigma= a>0,\, -\infty < \gamma < \infty \}\, .
\end{equation}

Thus, one can rewrite (13) as 
\begin{equation}
e^{tA} - e^{tB} = \int_0^{t} d\tau\, e^{(t-\tau)A}\,\left[A-B\right]\,e^{\tau B}\, .
\end{equation}

Per contra
\begin{equation} 
f(x,t) \equiv e^{tL}g(x)\, ,
\end{equation}
indeed, due to (9), (11)
\begin{equation}
\tilde{f}(x,s) \equiv \mathcal{L}\left\{ e^{tL} \right\}g(x) = R_s(L)g(x)\,.
\end{equation}

Finally, if $A=L(\beta)$ and $B=L_0$ by means of (16) - (17) we get the desired formula,
\begin{equation}
f(x,t) = e^{tL}g(x) = \bigg[ e^{tL_0} + \beta\int_0^{t} d\tau\, e^{(t-\tau)L}\,L_1 \,e^{\tau L_0} \bigg]g(x)\,
\end{equation}
or
\begin{equation}
\begin{aligned}
f(x,t) &= \sum^{+\infty}_{k=0} \beta^{k} \left[e^{tL_0}L_1 \ast\right]^{k}e^{tL_0}g(x)\\
&= \sum^{N}_{k=0} \beta^{k} \left[e^{tL_0}L_1 \ast\right]^{k}e^{tL_0}g(x) + o(\beta^{N})
\end{aligned}
\end{equation}
where
\begin{equation*}
(u\ast v)(t)=\int _{0}^{t}d\tau\,\, u(t-\tau)v(\tau )\,, 
\end{equation*}

i.e. at fixed perturbation $\beta$, by variation of $N$, we can reach a given accuracy.
In order to reach higher accuracies, the $exp(tL)$ in (19) should be calculated recursively 
substituting the equation itself under the integral.

Notably, for $N=1$
\begin{equation}
\begin{aligned}
f(x,t) &= e^{t(L_0 + \beta L_1)}g(x)\\
&=\bigg[ e^{tL_0} + \beta\int_0^{t} d\tau\, e^{(t-\tau)L_0}\, L_1\, e^{\tau L_0} +o(\beta) \bigg]g(x)\,. 
\end{aligned}
\end{equation}

Hence, the evolution for initial function $f(x,0) = g(x)$ is found. Then, 
after selecting the point(state) in the phase space, e.g. the maximum of $f(\,\cdot\,,t)$,
for each $t$ we pursue the time evolution of the point(state), thus determining 
the evolution of the Hamiltonian system in those time instances. 

Presented method can be slightly modified (by adding an extra variable $t$ with 
$\dot{t}=1$ equation) for the possible explicit time-dependent Hamiltonian systems.

The developed scheme, besides providing computation accuracy independent on time, also allows one to investigate the qualitative properties of the Hamiltonian system within the ergodic theory approach. Namely, despite the fact that analytical solution for Hamiltonian equations may not even exist due to internal resonances between different degrees of freedom, the precise time evolution for any state can be derived using the developed method. Such results can be used to distinguish to which class of dynamical systems the investigated system belongs to, e.g. of ergodic, weak mixing, K-system, etc. 

The obtained exact form of $f(\,\cdot\,,t)$ is more convenient for the study of asymptotic and average properties of the system. In a Hilbert space the time-average of the evolution (20) of the system is obtained by von Neumann's theorem as
\begin{equation}
\lim_{T\to\infty} \frac{1}{T} \int_{0}^T e^{t L} g(x) dt = \langle e^{tL}g(x), 1\rangle\,. 
\end{equation}

\section{Error-free evolution: sample systems}

The elimination of the explicit Laplace transform provides superior feasibility for computer algebra calculation 
than those described in \cite{GK1},\cite{GK2}. Note, that one is free to choose the initial function $g(x)$ 
from the set of the smooth functions upon the convenience for the studied system or the aims.

For the sample system, we will consider $\beta\ll 1$ for simplicity. It is important to note that, 
in the case of sufficiently large perturbations i.e. $\beta\to\infty$, by applying duality principle 
\cite{Fr}, one obtains rescaled Hamiltonian and by "Dual" KAM theorem perturbed invariant tori stay 
preserved.
 
Consider a dynamical system given by a two-dimensional Hamiltonian of two oscillators 
in action-angle form with different combination of a small perturbation
\begin{align}
H(I,\vartheta,\beta) &= \omega_1\, I_1 + \omega_2\, I_2 +\beta V(I,\vartheta)\,,\\
V_1(I,\vartheta)&=I_1 \cos\vartheta_2\,,\notag\\
V_2(I,\vartheta)&=I_1 \sin\vartheta_2+I_2 \cos\vartheta_1\notag\,.
\end{align}

We have the following exact form for the operator
\begin{equation}
e^{t L_0}g(I, \vartheta) = g(I,\vartheta+t \omega).
\end{equation}
The initial function $f(I, \vartheta, 0)$ was chosen in the form
\begin{equation}
g(I, \vartheta) = -\frac{1}{2}\sum_{k=1}^{n}\big(I_k-I_{k0}\big)^2+\sum_{k=1}^{n}\cos\big(\vartheta^i-\vartheta^i_0\big).
\end{equation}
 
Then the evolution of the systems by means of the described procedure is obtained respectively
\begin{equation}
\begin{array}{l}
f_1(I,\vartheta,t)=-\frac{1}{2}\Big((I_1-I_{10})^2+(I_2-I_{20})^2\Big)\\[5pt]
+\cos(\vartheta_1-\vartheta_{10}+t \omega_1)+\cos(\vartheta_2-\vartheta_{20}+t \omega_2)\\[5pt]
+\omega_2^{-1}\Big(\beta (I_1 (I_2-I_{20}) (\cos(\vartheta_2+t \omega_2)-\cos\vartheta_2)\\[5pt]
+(\sin\vartheta_2-\sin(\vartheta_2+t \omega_2)) \sin(\vartheta_1-\vartheta_{10}+t \omega_1))\Big)\,, 
\end{array}
\end{equation}
\begin{equation}
\begin{array}{l}
f_2(I,\vartheta,t)= (2\,\omega_1\omega_2)^{-1}
\bigg[2\beta\omega_1\Big((\cos(\vartheta_2 + t \omega_2)-\cos\vartheta_2)\\[5pt]
\times \sin(\vartheta_1 - \vartheta_{10} + t\omega_1) + 
I_1 (I_2 - I_{20}) (\sin(\vartheta_2 + t\omega_2) - \sin\vartheta_2)\Big) \\[5pt]
-\omega_2\Big(2\beta\Big((I_1-I_{10})I_2(\cos\vartheta_1-\cos(\vartheta_1 + t\omega_1)) + 
(\sin(\vartheta_1 + t\omega_1)\\[5pt]
-\sin\vartheta_1) \sin(\vartheta_2 - \vartheta_{20} + t \omega_2)\Big) + 
\omega_1\Big((I_1-I_{10})^2 + (I_2-I_{20})^2 \\[5pt]
-2\cos(\vartheta_1 - \vartheta_{10} + t \omega_1) - 
2\cos(\vartheta_2 - \vartheta_{20} + t \omega_2)\Big)\Big)\bigg]\,.
\end{array}
\end{equation}

\begin{figure}[t!]
\center{\includegraphics[width=1.1\linewidth]{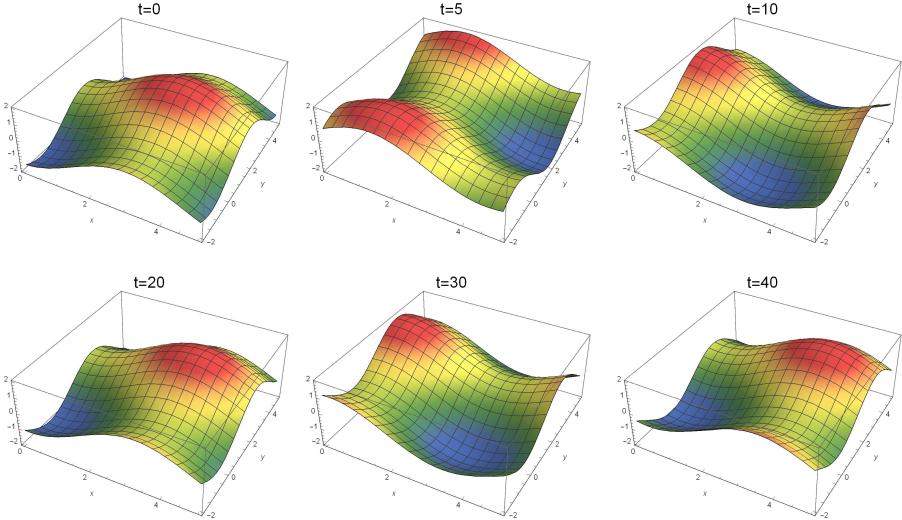}}
\caption{The time evolution of the surface determined by Eq. (26) for $I_1=const, I_2=const, \beta=0.03$.
The motion is regular with $T=20$ period.}
\end{figure}

\begin{figure*}[t!]
\center{\includegraphics[width=.7\linewidth]{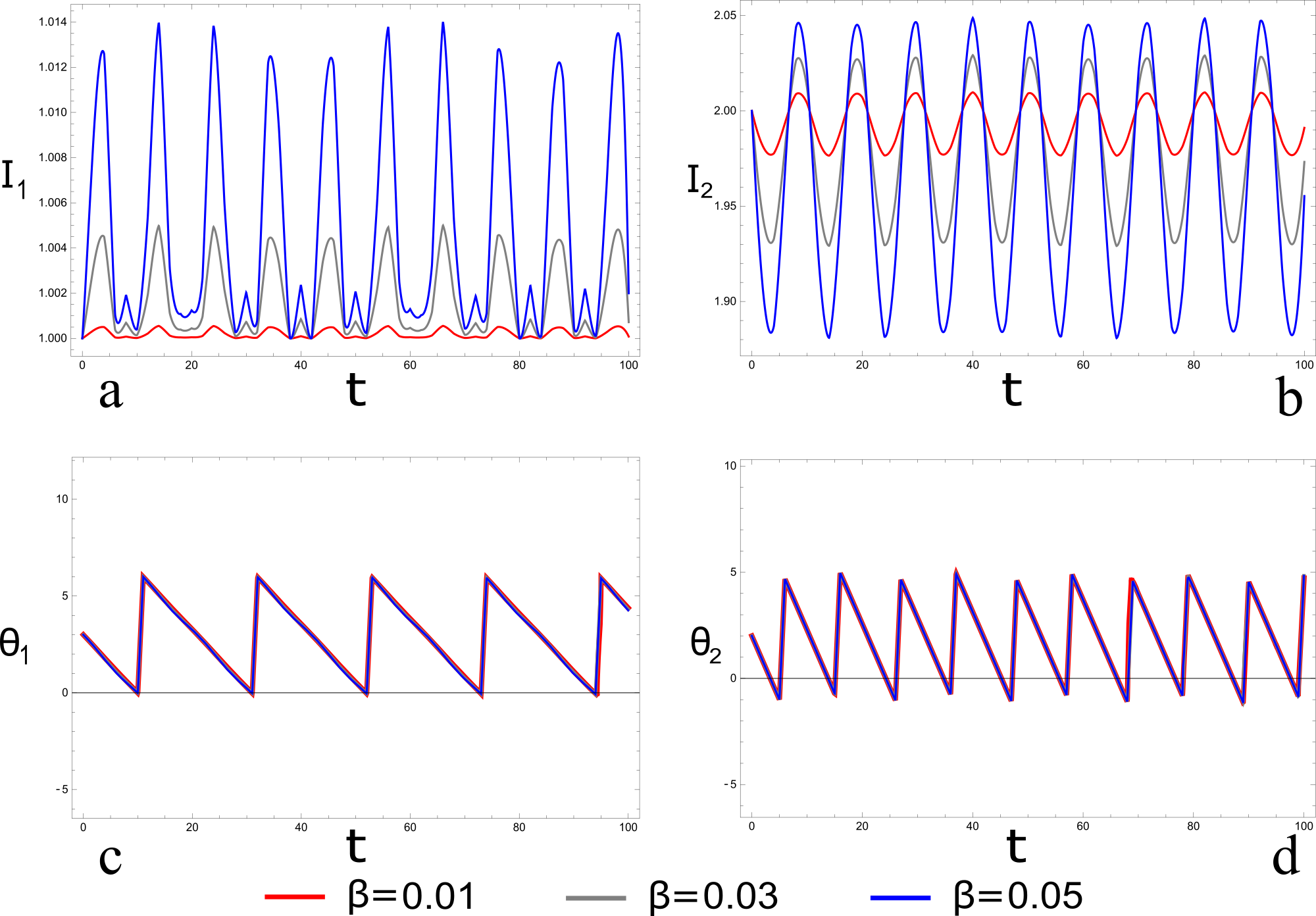}}
\caption{Variation in time of the system given by Eq. (26) 
with initial conditions $I_{10}=1.0, I_{20} = 2.0, \vartheta_{10}=3.0, 
\vartheta_{20}=2.0$ for different values of $\beta$.}
\end{figure*}

\begin{figure*}[t!]
\center{\includegraphics[width=.7\linewidth]{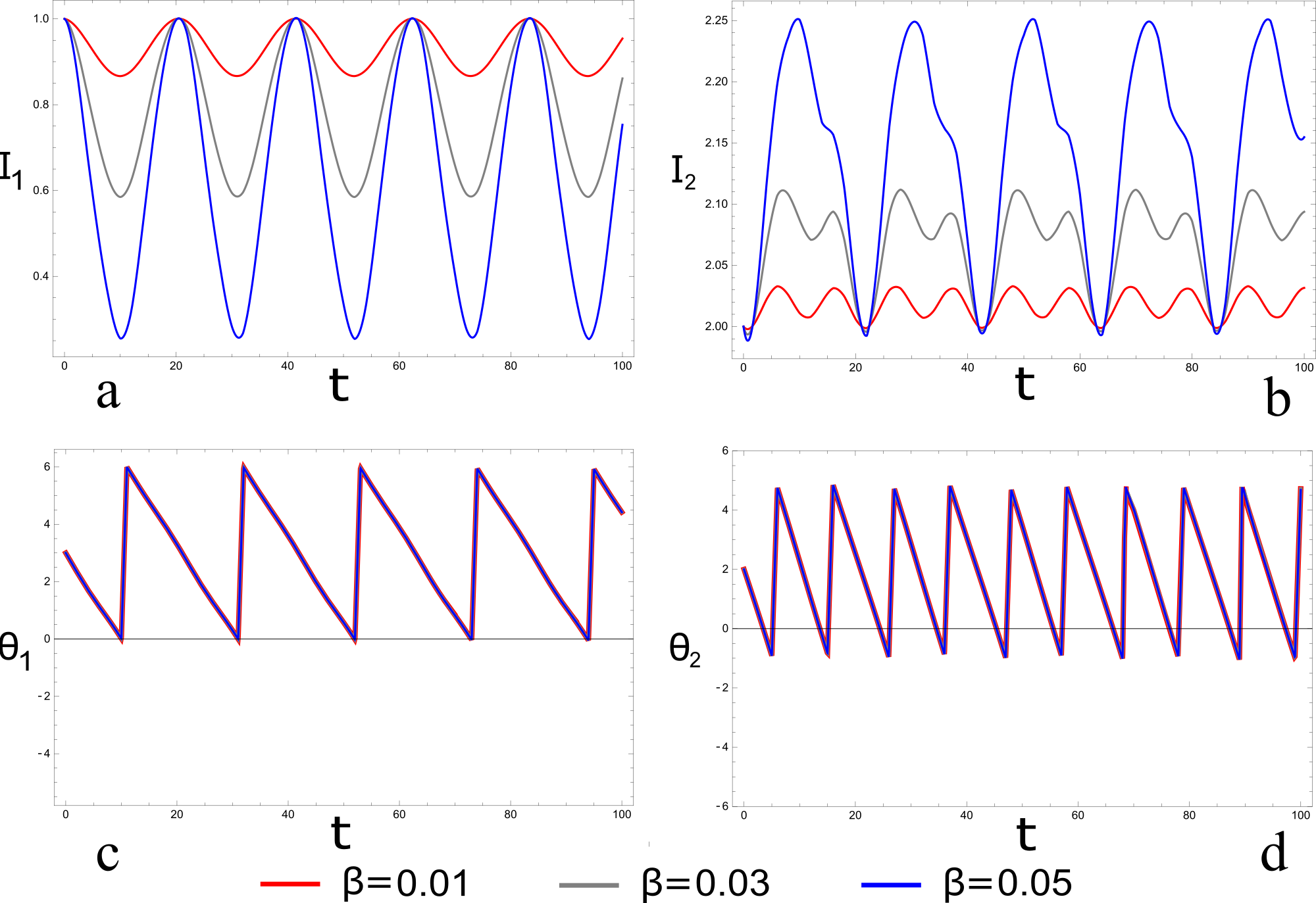}}
\caption{Variation in time of the system given by Eq. (27) 
with initial conditions $I_{10}=1.0, I_{20} = 2.0, \vartheta_{10}=3.0, 
\vartheta_{20}=2.0$ for different values of $\beta$.}
\end{figure*}

\begin{figure*}[t!]
\center{\includegraphics[width=1\linewidth]{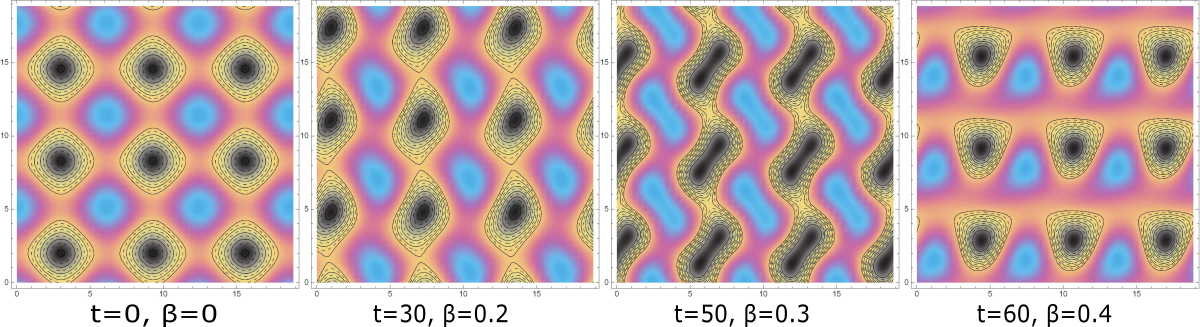}}
\caption{Evolving contour lines for $V_2(I,\vartheta)$ 
with increasing time and perturbation. The deformated tori at $t=60$ reappear again as stated by KAM theorem.}
\end{figure*}

The evolution of the surface determined by the function $f_1(I, \vartheta, t)$, i.e. corresponding to $V_1(I,\vartheta)$ 
perturbation system, at different instances of time for $I_1=const, I_2=const$ is shown in Fig. 1. The surface change 
and therefore the trajectories are regular in view of $t=0, 20, 50$ and $t=10, 30$ matching. 
In Figs. 2 and 3 the variations of $(I,\vartheta)$ by the time for different values of $\beta$ are shown. 
One can observe deviations compared with the unperturbed (integrable) system's evolution. 
In Fig. 4 the contour lines are presented for the function $f_2(I, \vartheta, t)$. The phase tori deformed at the 
increase of $\beta$ in time are not destroyed, but remain only slightly deformed, as stated by KAM theory.

During computations, at the increase (decrease) of  $\beta$  by an order of magnitude,  the calculation time varied proportional, at the accuracy $10^{-6}$.

\section{Conclusions}

We described a method of numerical investigation of evolution of dynamical systems independent on time accuracy, i.e.
not based on iterations as the typical numerical codes but based on exact computation of the evolution function. In such
formulation the method enables the use of computer algebra.

Essential point here is that the avoidance of the Laplace transform and spatial integration eliminates possible imaginary component appearance 
during computer algebra calculation, as could happen at the previous procedure. Then the resulting form for the evolution 
function is far more practical in further study and analysis. We demonstrated the method on an example of two-dimensional Hamiltonian of two oscillators and the error-free time evolution was shown for various values of the perturbation parameter $\beta$; the survival of the
perturbed tori was shown in accordance to KAM theory.   

The developed method can be applied to various physical and astrophysical many-dimensional non-linear systems, most notably, to gravitating N-body problem, i.e. to the long term planetary system evolution, galaxy, galaxy cluster dynamics.  

Another intriguing area of application can be the evolution of cosmological perturbations vs N-body formalism, particularly, for scalar perturbations within the discrete gravitating masses approach of \cite{Ein1} due to the possibility of adopting the modified method (dependent on the specific prospect) to dissipative systems (e.g. in quasistatic approximation or via scheme discussed in \cite{Bat}). Then, after estimation of the error-free expressions, the cosmic structure formation and evolution, back reaction effects can be revealed in more details.

\begin{acknowledgements}
We are grateful to the referee for valuable suggestions and comments.
\end{acknowledgements}

\end{document}